\documentclass{aa}

\usepackage{txfonts}
\usepackage{graphicx}
\usepackage{mathtools}
\usepackage{float}
\usepackage{epstopdf}
\usepackage{amsmath}

\bibpunct{(}{)}{;}{a}{}{,} % to follow the A&A style

\begin{document}

\title{Conditions for electron-cyclotron maser emission in the solar corona}

%----------------- Authors ---------------

\author{D. E. Morosan\inst{\ref{1}}
 \and 
P. Zucca\inst{\ref{2}}
 \and 
D. S. Bloomfield\inst{\ref{1}}
 \and
P. T. Gallagher\inst{\ref{1}} 
}

%----------------- Institutes ---------------
\institute{ School of Physics, Trinity College Dublin, Dublin 2, Ireland \label{1}
\and
LESIA, UMR CNRS 8109, Observatoire de Paris, 92195  Meudon, France \label{2}
}

\date{ Received /
		Accepted }

\abstract{The Sun is an active source of radio emission ranging from long duration radio bursts associated with solar flares and coronal mass ejections to more complex, short duration radio bursts such as solar S bursts, radio spikes and fibre bursts. While plasma emission is thought to be the dominant emission mechanism for most radio bursts, the electron-cyclotron maser (ECM) mechanism may be responsible for more complex, short-duration bursts as well as fine structures associated with long-duration bursts.}
{We investigate the conditions for ECM in the solar corona by considering the ratio of the electron plasma frequency $\omega_p$ to the electron-cyclotron frequency $\Omega_e$. The ECM is theoretically possible when $\omega_p/\Omega_e~<~1.$}
{Two-dimensional electron density, magnetic field, plasma frequency, and electron cyclotron frequency maps of the off-limb corona were created using observations from SDO/AIA and SOHO/LASCO, together with potential field extrapolations of the magnetic field. These maps were then used to calculate $\omega_p$/$\Omega_e$ and Alfv\'en velocity maps of the off-limb corona.}
{We found that the condition for ECM emission ($\omega_p/\Omega_e<1$) is possible at heights $<1.07~R_\sun$ in an active region near the limb; that is, where magnetic field strengths are $>40$~G and electron densities are $>3\times10^8$~cm$^{-3}$. In addition, we found comparatively high Alfv\'en velocities ($>0.02$~$c$ or $>6000$~km~s$^{-1}$) at heights $<1.07~$R$_\sun$ within the active region.}
{This demonstrates that the condition for ECM emission is satisfied within areas of the corona containing large magnetic fields, such as the core of a large active region. Therefore, ECM could be a possible emission mechanism for high-frequency radio and microwave bursts. }


\maketitle

\section{Introduction}

{Solar activity is often accompanied by solar radio emission, such as Type I--V bursts, in addition to a variety of more complex short timescale bursts. Some examples are zebra patterns \citep{sl72}, radio spikes \citep{be82, be96}, fibre bursts \citep{ra08}, and solar S bursts \citep{mcc82, mo15}. Electron-cyclotron maser (ECM) emission has been proposed as one of the possible emission mechanisms for some of these radio bursts.}

{The requirement for ECM to generate emission is that the electron plasma frequency ($\omega_p$) is less than the electron-cyclotron frequency ($\Omega_e$) at the emission site \citep{mel91}. In planetary magnetospheres, ECM is well established as the emission mechanism of some radio bursts, as the condition $\omega_p < \Omega_e$ can be easily satisfied. For example, the auroral kilometric radiation (AKR) of the Earth was found to be generated by ECM in electron density depletion regions from 1.5--2.5~R$_{\oplus}$, where $\omega_p/\Omega_e = 0.14$ and the Alfv\'en speed is $v_A = 0.17~c$ in small localised regions of only a few kilometres in size observed \textit{in situ} by the Viking Satellite \citep{hi92} .}

{ECM emission was also proposed in solar physics for the interpretation of the coherent emission mechanism of Type IV bursts \citep{wa04}, fine structure in Type IV bursts \citep{as88}, and zebra patterns that also often accompany Type IV bursts \citep{tr11} to explain the high brightness temperature and high degree of polarisation of these bursts. \citet{ta13} extended this theory to account for the characteristics of Type V radio bursts as well.}

{Spike bursts in the solar corona are assumed to be generated by ECM on the basis of an analogy with the AKR of the Earth \citep{mel94}. Spike bursts are seen during the impulsive phase of solar flares and they appear as short, narrowband spikes near the starting frequency of Type III bursts. \citet{mel94} suggested that the requirement of $ \omega_p < \Omega_e$ for ECM is not satisfied near the energy release site in the corona where Type III radio bursts may originate as plasma emission, but in neighbouring low-density regions where it is more likely to obtain $ \omega_p < \Omega_e$. }

\begin{figure*}[t]
\sidecaption
\includegraphics[angle = 90, width = 365px, trim = 160px 10px 140px 20px ]{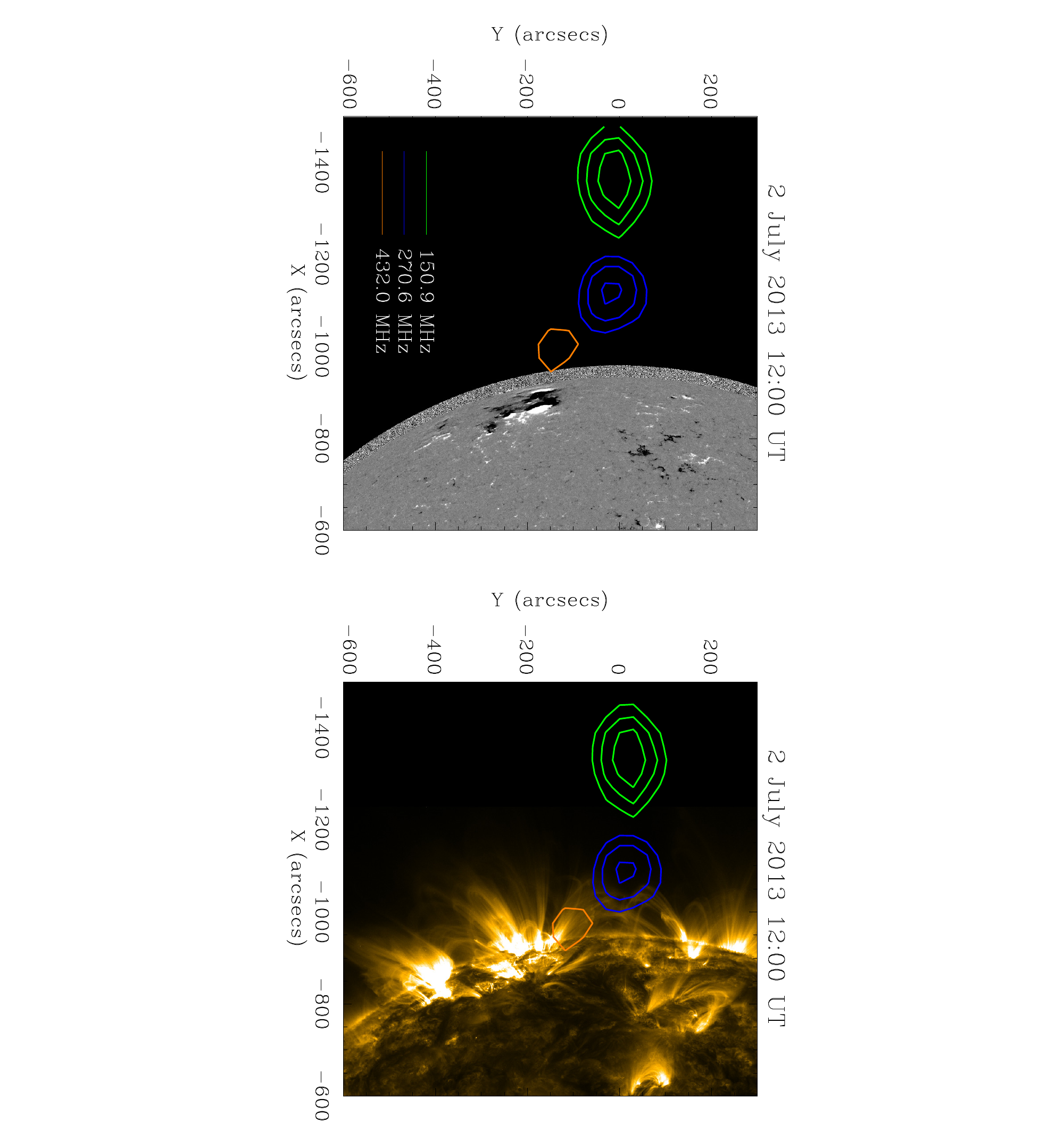}
\caption{Active region NOAA 11785 when it is visible near the solar limb on 2 July 2013  in the HMI magnetogram (left panel) and in the AIA 171~$\AA$ images (right panel) superimposed with the Type I storm, which was visible from 2 July onwards, observed by the Nan{\c c}ay Radioheliograph at different frequencies.  \label{fig1}}
\end{figure*}

{Solar S bursts are very similar in appearance to Jovian S bursts that are a well-accepted example of ECM emission. They have very narrow bandwidths and short lifetimes and are significantly more polarised than Type III radio bursts, and this is indicative of ECM emission \citep{mo15}. Jovian S bursts are generated by $\sim$5~keV electrons accelerated in the flux tubes connecting Io to Jupiter \citep{za96}. These electrons have an adiabatic motion along the magnetic field lines and are magnetically mirrored at the foot of the Io flux tube. Near Jupiter, they emit radio bursts triggered by the loss cone instability in the magnetically mirrored population of electrons. Simulations show that these electrons are accelerated by Alfv\'en waves in a very narrow region named the Alfv\'en wave resonator where short-lived electron beams are produced \citep{he07, he09}.}

{\citet{re15} took a new approach to investigate whether ECM is a viable emission mechanism in the solar corona by considering the ratio $\omega_p$/$\Omega_e$. A hydrostatic density model was used to estimate the value of electron density, and, hence $\omega_p$, while potential and non-linear force-free models were used to estimate the magnetic field, and hence $\Omega_e$. In the case of four active regions studied, it was found that the smallest values of $\omega_p$/$\Omega_e$ are located where the magnetic field strength is the largest at the bottom of the corona directly above sunspots at heights $<1.2$~R$_\sun$.}

{In this Letter, we explore the ratio of $\omega_p$/$\Omega_e$ using data-constrained density and magnetic field models of the solar corona in order to determine if ECM is a viable emission mechanism for solar radio bursts. }

\section{The electron-cyclotron maser}

{The requirement for ECM to generate radio emission is that $\omega_p$ is less than $\Omega_e$ at the emission site \citep{mel91}. These frequencies are given by,
\begin{equation}
\omega_p = \sqrt{\frac{n_e e^2}{m_e\epsilon_0}}~~ ,
\end{equation}
and,
\begin{equation}
\Omega_e= \frac{eB}{m_e}~~ ,
\end{equation} 
where $n_e$ is the electron plasma density, $B$ is the magnetic field strength and the remaining quantities are known physical constants. }

{The ratio $\omega_p/\Omega_e$ is also related to the Alfv\'en speed,
\begin{equation}
v_A = \frac{B}{\sqrt{\mu_0 n_i m_i}}~~,
\end{equation}
where $n_i$ is the ion density and $m_i$ the ion mass. The ratio $\omega_p/\Omega_e$ can then be expressed in the following way by combining Equations (1), (2) and (3) and assuming $n_i=n_e$,
\begin{equation}
\frac{\omega_p}{\Omega_e} = \sqrt{\frac{m_e}{m_i}}\frac{c}{v_A}~~ .
\end{equation}
In order for $\omega_p/\Omega_e$ to be less than unity and hence ECM to occur \citep{mel91}, $v_A$ needs to be $ > 0.02~c$ ($>6000$~km~s$^{-1}$) assuming that the mean molecular weight, $\mu$, is 0.6 in the corona. Such Alfv\'en speeds are high by comparison with normal coronal conditions, although not unexpected in active regions with high magnetic field strengths \citep{wa05}.}

\section{Results}

{In order to satisfy the condition $ \omega_p < \Omega_e$ in the solar corona, a high $B$ and a low $n_e$ is required at the emission site. Therefore, we investigated the conditions in the corona in the vicinity of the $\beta\gamma\delta$ active region NOAA 11785 on 1 July 2013  when it was located on the eastern solar limb.  Figure 1 shows the active region on 2 July 2013 when it is visible on the solar disc near the eastern solar limb in the Heliospheric and Magnetic Imager (HMI) magnetogram (left panel) and in the Atmospheric Imaging Assembly  \citep[AIA;][]{le12} 171~$\AA$ image (right panel) both of which are instruments onboard the Solar Dynamics Observatory (SDO). Superimposed on these are the contours of the Type I storm observed by Nan{\c c}ay Radioheliograph at three different radio frequencies. We chose this active region since it was associated with significant radio activity during its rotation: a long-lived Type I noise storm (Figure 1); Type III and Type IIIb radio bursts; and solar S bursts \citep{mo15}. }

\begin{figure*}[ht]
\centering
\includegraphics[width = 530px, trim = 0px 70px 0px 70px ]{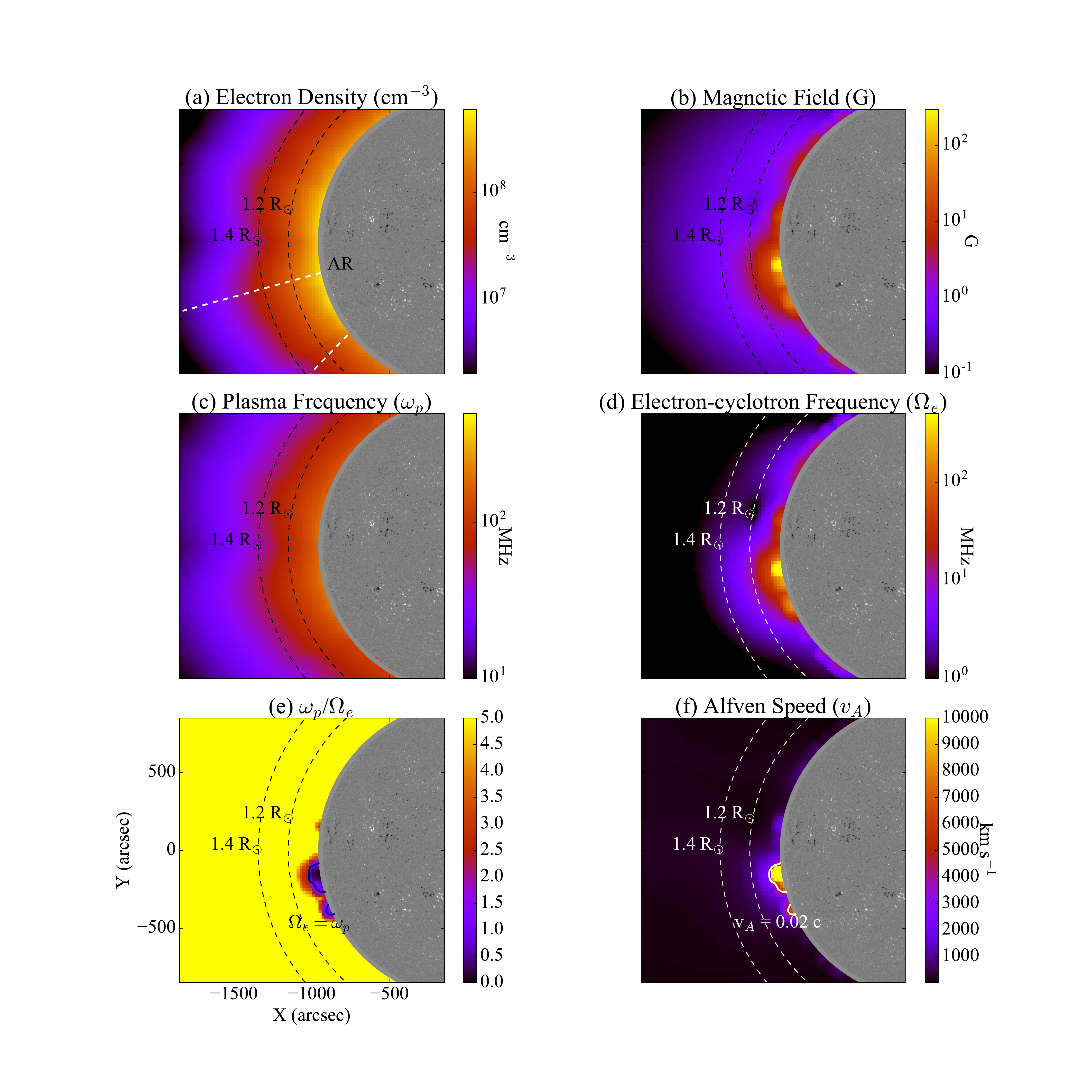}
\caption{Characteristics of the solar corona on 1 July 2013 when the $\beta\gamma\delta$ active region was located on the solar limb: (a) electron density, (b) magnetic field, (c) plasma frequency, (d) electron-cyclotron frequency, (e) plasma frequency to electron-cyclotron frequency ratio and, (f) Alfv\'en speed. The location of the active region on the limb (AR) is labelled in panel (a). \label{fig2}}
\end{figure*} 

{We constructed a map of the electron densities to determine $\omega_p$ (Equation 1) and a map of the magnetic field strength to determine $\Omega_e$ (Equation 2). We used the method of calculating the electron densities of \citet{zu14} on 1 July 2013 when NOAA 11785 was  located on the eastern solar limb (Figure 2a) so that the plane-of-sky densities are centred on the Carrington longitude of the active region. Electron densities in the corona were estimated in two ways: for the height range 1--1.3~R$_\sun$, densities were obtained from the differential emission measure derived from the six coronal filters of AIA and, for the height range 2.5--5~R$_\sun$ they were obtained using polarised brightness from the Large Angle and Spectrometric Coronograph \citep[LASCO;][]{br95} on-board the Solar and Heliospheric Observatory (SOHO). For the intermediate height range 1.3--2.5~R$_\sun$, a combined plane-parallel and spherically symmetric model was employed (see \citealt{zu14} for further details).} 

{The magnetic field in Figure 2b was estimated using a potential field source surface (PFSS) model that provides an approximation of the coronal magnetic field at heights up to 2.5~R$_\sun$ based on the observed photospheric field \citep{sc03}. We used the PFSS solution from 4 July 2013 when the active region was not as close to the limb to extract the longitudinal slice through the active region that is equivalent to the Earth-viewed plane-of-sky on 1 July 2013. This was necessary as the photospheric magnetic field is only accurately represented in the PFSS bottom boundary when the active region is not located close to the east limb (owing to the transition from the flux-transported surface field on the unobserved hemisphere to the observed field on the Earth-facing hemisphere). In addition, the magnetic field structures above the Carrington longitude of the active region in Figure 2b remain consistent throughout a period of seven consecutive days thereafter, so no major changes occurred.}

{Using Equations (1), (2) and, (4), we created maps of $\omega_p$ (Figure 2c), $\Omega_e$ (Figure 2d), the ratio $\omega_p/\Omega_e$ (Figure 2e) and, $v_A$ (Figure 2f). Figure 2e shows that the locations where $\omega_p < \Omega_e$ occurs are at heights below $1.1~R_\sun$ in the active region. In addition, high Alfv\'en velocities are present at the same locations in Figure 2f ($>0.02~c$ or $>6000$~km~s$^{-1}$). We can therefore conclude that ECM emission is a possible emission mechanism in the solar corona, but only at heights  below 1.1~R$_\sun$ in the active region studied. These results agree with \citet{re15}, where estimates of the $\omega_p/\Omega_e$ ratio showed that ECM is possible at heights up to 1.2~R$_\sun$ based on a sample of four active regions.}

{The radial profiles of $n_e$, $B$, $\omega_p$, $\Omega_e$, $\omega_p/\Omega_e$ and $v_A$ are shown in Figure 3 using the same panel labelling as in Figure 2. These plots are taken along the two dashed lines in Figure 2a, one for the active region (red solid line in Figure 3) and one for a quiet Sun region (purple dashed line). The quiet Sun magnetic field is too low to allow for ECM emission and the quiet Sun density is not low enough to compensate for this lack of high field strength. A high magnetic field strength is the most important factor in determining whether $\omega_p < \Omega_e$ even inside an active region where densities are considered to be high. In Figure 3e, the shaded region shows that maser emission is possible up to a height of 1.07~R$_\sun$. As a result, ECM emission can only occur in a small area within the active region where the magnetic field is highest. }

\begin{figure}[t]
\centering
\includegraphics[width = 260px, trim = 20px 60px 20px 30px ]{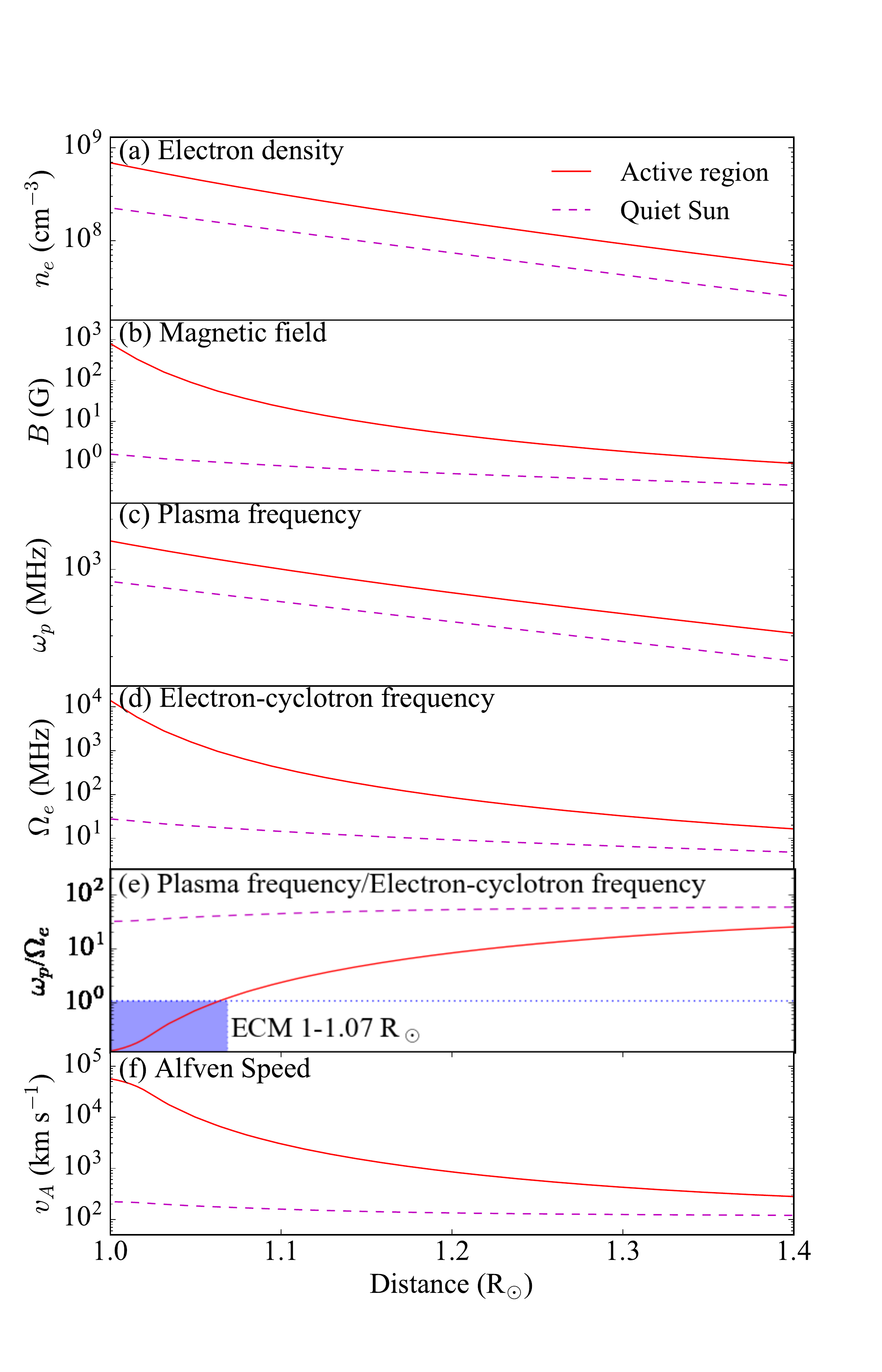}
\caption{ Radial profiles of (a) electron density, (b) magnetic field, (c) plasma frequency, (d) electron-cyclotron frequency, (e) ratio of the two frequencies and (f) Alfv\'en speed along the dashed line in Figure 2a passing through the active region (red solid line) and the dashed line passing through the quiet Sun in Figure 2a (purple dashed line). In panel (e), the blue dotted line indicates the $\omega_p/\Omega_e = 1$ level while the blue shaded box represents the height range in the corona where ECM emission is possible.  \label{fig3}}
\end{figure} 

\section{Conclusions}

{In this Letter, we investigated the possibility of ECM emission in the solar corona as a viable emission mechanism for solar radio bursts. We found that on 1 July 2013, when a $\beta\gamma\delta$ active region associated with significant radio activity was on the solar limb, the condition $\omega_p < \Omega_e$ necessary for ECM was met within the active region at a height up to 1.07~R$_\sun$. We also observe high Alfv\'en velocities at those heights, as a result of the high concentration of strong magnetic field in the active region. }

{The conditions for ECM in the solar corona were met at extremely low heights that do not coincide with heights at which solar S bursts have been observed at low frequencies \citep{mo15}, but at a height appropriate to where high-frequency radio and microwave emission (>500~MHz) can occur (radio spikes, Type IV bursts and associated fine structures). In this study we only considered PFSS magnetic field models. \citet{re15} also considered non-linear force-free (NLFF) models and showed that the area and height where ECM conditions are favourable increase when NLFF models are used. However, the maximum height found was also in the low corona ($\sim$1.2~R$_\sun$).

{Previous studies have suggested that values of $\omega_p/\Omega_e<1$ can be found higher in the corona, either where highly twisted flux tubes exist in the magnetic configuration \citep{re15} or density depleted ducts are found \citep{wa15}. An erupting flux rope would also create ECM favourable conditions at greater heights provided that the magnetic field inside is $>10$~G for densities $>10^7$~cm$^{-3}$ that are found up to $\sim$2~R$_\sun$. However, such high magnetic field values have not been observed. \citet{wa15} estimated that if Type III radio bursts were generated by maser emission they would escape along density depleted ducts and we would observe fundamental-harmonic pairs when $0.1< \omega_p/\Omega_e<0.4$ and harmonic bursts when $0.4< \omega_p/\Omega_e<1.4$. However, maser generated Type IIIs would only escape the corona along density depleted ducts, while observations suggest they are directly accelerated in active regions \citep{ch13} and plasma emission is the more widely accepted mechanism for Type III bursts. }

{High cadence interferometric images ($<$0.1~s) are therefore needed for imaging of the fine structures in radio bursts at various frequencies. Their locations can then be compared to the background electron density and magnetic field in the corona to better understand their emission mechanism. Currently, LOFAR is the only instrument capable of such observations at low radio frequencies (<240~MHz) and has already provided further insight into numerous poorly studied radio bursts \citep{mo14,mo15}. At high radio frequencies, future solar dedicated instruments such as eSolar will be  necessary for the study of solar radio emission and plasma processes in the low corona.}

\begin{acknowledgements}{This work has been supported by a Government of Ireland studentship from the Irish Research Council (IRC), the IRC New Foundations Award and the Non-Foundation Scholarship awarded by Trinity College Dublin. DSB received funding from the European Space Agency PRODEX Programme and the European Union's Horizon 2020 Research and Innovation Programme under grant agreement No. 640216 (FLARECAST project). We would like to thank Hamish Reid and Philippe Zarka for stimulating conversations on the topic of ECM.}\end{acknowledgements}

% for the bibliography, at the end
\bibliographystyle{aa} % style aa.bst
\bibliography{Morosan_letter.bib} % your references file.bib

\end{document}